\documentclass[12pt]{article}

\input{epsf.tex}

\newcommand{\E}{{\cal{E}}}

\renewcommand{\a}{\alpha}
\renewcommand{\d}{{\rm d}}
\newcommand{\dfrac}[2]{\displaystyle\frac{#1}{#2}}
\newcommand{\be}{\begin{equation}}
\newcommand{\ee}{\end{equation}}
\newcommand{\bea}{\begin{eqnarray}}
\newcommand{\eea}{\end{eqnarray}}

\begin{document}
\title{\bf\Large The double--Kerr equilibrium configurations involving one extreme
object}
\author{J A Rueda$^1$, V~S~Manko$^2$, E Ruiz$^3$ and J D Sanabria--G\'omez$^1$}
\date{}
\maketitle

\vspace{-1cm}

\begin{center}
$^1$ Escuela de F\'{i}sica, Universidad Industrial de Santander,
A.A. 678, Bucaramanga, Colombia\\
$^2$ Departamento de F\'\i sica, Centro de Investigaci\'on y de
Estudios Avanzados del IPN, A.P. 14-740, \\07000 M\'exico D.F.,
Mexico\\
$^3$ Area de F\'\i sica Te\'orica, Universidad de Salamanca,
\\37008 Salamanca, Spain
\medskip
\end{center}

\vspace{.1cm}

\begin{abstract}
\noindent We demonstrate the existence of equilibrium states in
the limiting cases of the double--Kerr solution when one of the
constituents is an extreme object. In the `extreme--subextreme'
case the negative mass of one of the constituents is required for
the balance, whereas in the `extreme--superextreme' equilibrium
configurations both Kerr particles may have positive masses. We
also show that the well--known relation $|J|=M^2$ between the mass
and angular momentum in the extreme single Kerr solution ceases to
be a characteristic property of the extreme Kerr particle in a
binary system.
\end{abstract}

\noindent \hspace{1cm} PACS number: 04.20.Jb

\vspace{3cm}

\section{Introduction}

In the paper \cite{MRS} a unified approach to the solution of the
double--Kerr equilibrium problem \cite{KNe} in its extended form
involving arbitrary combinations of the sub- and superextreme Kerr
particles was developed. This permitted later on to rigorously
prove the non--existence of equilibrium states of two Kerr black
holes with positive masses \cite{MRu1} and, furthermore, to
establish the general law determining equilibrium of two arbitrary
Kerr particles \cite{MRu2}. In the paper \cite{MRS} the explicit
formulae defining the double--Kerr solution in the limiting case
when one of the constituents is an extreme object and the other a
sub- or superextreme Kerr particle were also worked out. At the
same time, attempts to solve numerically the corresponding balance
equations failed to result in any equilibrium configuration, which
forced the authors of \cite{MRS} to put forward a conjecture on
the impossibility of balance between an extreme and a non--extreme
Kerr particles via the gravitational spin--spin interaction.

However, a recent approximation analysis of the double--Kerr
equilibrium problem carried out in \cite{MRu3}, especially the
last approximation scheme in which, on the one hand, the upper
constituent looks very much like an extreme object and, on the
other hand, the corresponding binary equilibrium configurations do
exist, strongly motivated us to undertake a revision of the
limiting case from \cite{MRS} involving one extreme Kerr particle
and thus continue search for the equilibrium configurations which
were not found before. As a result, we have been able to arrive,
after scrupulously modifying our schemes for finding numerical
roots of the analytical system of balance equations, at the
desired equilibrium configurations which can be considered as the
first examples of the balance of that kind in the literature. As
expected, the equilibrium states between one extreme and one
subextreme Kerr particles (the `extreme--subextreme'
configurations) necessarily imply the negative mass of one of the
constituents; in the `extreme--superextreme' case the equilibrium
configurations are possible when both Kerr particles possess
positive Komar masses.

The rest of the paper is organized as follows. In section~2 we
write down the metric describing the stationary axisymmmetric
double--Kerr systems with one extreme constituent, as well as the
corresponding balance conditions. Particular equilibrium states
are considered in section~3 and some of their physical
characteristics such as Komar masses and angular momenta
\cite{Kom} are calculated. Section~4 is devoted to the discussion
of the results obtained.

%%%%%%%%%%%%%%%%%%%%%%%%%%%%%%%%%%%%%%%%%%%%%%%%%%%%%%%%%%%%%%
\section{The `extreme--non-extreme' systems and the
balance conditions}
%%%%%%%%%%%%%%%%%%%%%%%%%%%%%%%%%%%%%%%%%%%%%%%%%%%%%%%%%%%%%%

For tackling the two--Kerr configurations with one extreme
particle we shall make use of the solution obtained in \cite{MRS}
with the aid of Sibgatullin's method \cite{Sib,MSi} which is
defined by the Ernst complex potential $\E$ \cite{Ern} of the
form\footnote{The wrong sign of $B$ in \cite{MRS} is rectified.}
$$ \E=\frac{\Lambda + \Gamma}{\Lambda - \Gamma}, $$ $$
\Lambda=\left|\begin{array}{cccc} \dfrac{r_1}{\alpha_1 - \beta_1}
& \dfrac{r_2}{\alpha_2 - \beta_1} & \dfrac{r_3}{\alpha_3 -
\beta_1}
& - \dfrac{r_1^2}{(\alpha_1 - \beta_1)^2}\vspace{0.25cm}\\
\vspace{0.25cm}\dfrac{r_1}{\alpha_1 - \beta_2} &
\dfrac{r_2}{\alpha_2 - \beta_2} & \dfrac{r_3}{\alpha_3 - \beta_2}
& - \dfrac{r_1^2}{(\alpha_1 - \beta_2)^2}\\
\vspace{0.25cm}\dfrac{1}{\alpha_1 - \bar\beta_1} &
\dfrac{1}{\alpha_2 - \bar\beta_1} & \dfrac{1}{\alpha_3 -
\bar\beta_1} & r_1^2\dfrac{\partial} {\partial
\alpha_1}\left[\dfrac{1}
{(\alpha_1-\bar\beta_1)r_1}\right]\\
\dfrac{1}{\alpha_1 - \bar\beta_2} & \dfrac{1}{\alpha_2 -
\bar\beta_2} & \dfrac{1}{\alpha_3 - \bar\beta_2} &
r_1^2\dfrac{\partial} {\partial \alpha_1}\left[\dfrac{1}
{(\alpha_1-\bar\beta_2)r_1}\right]\\
\end{array}\right|, $$ $$
\Gamma=\left|\begin{array}{ccccc} 0 & 1 & 1 & 1 &
(z-\a_1)/r_1\vspace{0.15cm}\\\vspace{0.25cm} 1 &
\dfrac{r_1}{\alpha_1 - \beta_1} & \dfrac{r_2}{\alpha_2 - \beta_1}
& \dfrac{r_3}{\alpha_3 - \beta_1} & - \dfrac{r_1^2}{(\alpha_1 -
\beta_1)^2}\\\vspace{0.25cm} 1 & \dfrac{r_1}{\alpha_1 - \beta_2} &
\dfrac{r_2}{\alpha_2 - \beta_2} & \dfrac{r_3}{\alpha_3 - \beta_2}
& - \dfrac{r_1^2}{(\alpha_1 - \beta_2)^2}\\\vspace{0.25cm} 0 &
\dfrac{1}{\alpha_1 - \bar\beta_1} & \dfrac{1}{\alpha_2 -
\bar\beta_1} & \dfrac{1}{\alpha_3 - \bar\beta_1} &
r_1^2\dfrac{\partial} {\partial \alpha_1}\left[\dfrac{1}
{(\alpha_1-\bar\beta_1)r_1}\right]\\
0 & \dfrac{1}{\alpha_1 - \bar\beta_2} & \dfrac{1}{\alpha_2 -
\bar\beta_2} & \dfrac{1}{\alpha_3 - \bar\beta_2} &
r_1^2\dfrac{\partial}{\partial \alpha_1}
\left[\dfrac{1}{(\alpha_1-\bar\beta_2)r_1}\right]\\
\end{array}\right|, $$ \be r_n =\sqrt{\rho^2 + (z-\alpha_n)^2},
\label{ernst_pot} \ee where $\a_1$ is an arbitrary real constant,
the constants $\a_2$ and $\a_3$ can assume arbitrary real values
or be an arbitrary complex conjugate pair $\a_3=\bar\a_2$, a bar
over a symbol denoting complex conjugation; $\beta_1$ and
$\beta_2$ are arbitrary complex parameters; $\rho$ and $z$ are the
usual Weyl--Papapetrou cylindrical coordinates in the Papapetrou
stationary axisymmetric line element
\begin{equation}\label{papa}
\d s^2= f^{-1}[e^{2\gamma} (\d\rho^2 + \d z^2) + \rho^2
\d\varphi^2]- f( \d t - \omega \d\varphi)^2.
\end{equation}

The metric functions $f$, $\gamma$ and $\omega$ in (\ref{papa})
corresponding to the potential (\ref{ernst_pot}) were also
calculated in \cite{MRS}, and these are given by the expressions
$$ f=\frac{\Lambda \bar\Lambda - \Gamma \bar\Gamma}{(\Lambda -
\Gamma)(\bar\Lambda - \bar\Gamma)}, \quad e^{2\gamma} =
\frac{\Lambda \bar\Lambda - \Gamma \bar\Gamma}{K_0\bar
K_0r_1^4r_2r_3} , \quad \omega = -\frac{2\,{\rm Im}[H(\bar\Lambda
- \bar\Gamma)]}{\Lambda\bar\Lambda - \Gamma\bar\Gamma}, $$ \bea
H=\left|\begin{array}{rcccc} 0 & g(\alpha_1) & g(\alpha_2) &
g(\alpha_3) & r_1^2\dfrac{\partial}
{\partial \alpha_1}\left( \dfrac{\alpha_1-z}{r_1} \right) \vspace{0.15cm}\\
-1 & \dfrac{r_1}{\alpha_1 - \beta_1} & \dfrac{r_2}{\alpha_2 -
\beta_1} & \dfrac{r_3}{\alpha_3 - \beta_1} & -
\dfrac{r_1^2}{(\alpha_1 -
\beta_1)^2} \vspace{0.25cm}\\
-1 & \dfrac{r_1}{\alpha_1 - \beta_2} & \dfrac{r_2}{\alpha_2 -
\beta_2} & \dfrac{r_3}{\alpha_3 - \beta_2} & -
\dfrac{r_1^2}{(\alpha_1 - \beta_2)^2} \vspace{0.25cm} \\
0 & \dfrac{1}{\alpha_1 - \bar\beta_1} & \dfrac{1}{\alpha_2 -
\bar\beta_1} & \dfrac{1}{\alpha_3 - \bar\beta_1} &
r_1^2\dfrac{\partial} {\partial \alpha_1}\left[\frac{1}
{(\alpha_1-\bar\beta_1)r_1}\right] \vspace{0.25cm}\\
0 & \dfrac{1}{\alpha_1 - \bar\beta_2} & \dfrac{1}{\alpha_2 -
\bar\beta_2} & \dfrac{1}{\alpha_3 - \bar\beta_2} &
r_1^2\dfrac{\partial}
{\partial \alpha_1}\left[\dfrac{1}{(\alpha_1-\bar\beta_2)r_1}\right]\\
\end{array}\right|, \nonumber \eea \bea
&&K_0 = \left|\begin{array}{cccc} \dfrac{1}{\alpha_1 - \beta_1} &
\dfrac{1}{\alpha_2 - \beta_1} & \dfrac{1}{\alpha_3 - \beta_1} &
\dfrac{1}{(\alpha_1 -
\beta_1)^2} \vspace{0.15cm}\\
\dfrac{1}{\alpha_1 - \beta_2} & \dfrac{1}{\alpha_2 - \beta_2} &
\dfrac{1}{\alpha_3 - \beta_2} &
\dfrac{1}{(\alpha_1 - \beta_2)^2} \vspace{0.15cm}\\
\dfrac{1}{\alpha_1 - \bar\beta_1} & \dfrac{1}{\alpha_2 -
\bar\beta_1} & \dfrac{1}{\alpha_3 - \bar\beta_1} &
\dfrac{1}{(\alpha_1-\bar\beta_1)^2}
\vspace{0.15cm}\\
\dfrac{1}{\alpha_1 - \bar\beta_2} & \dfrac{1}{\alpha_2 -
\bar\beta_2} &
\dfrac{1}{\alpha_3 - \bar\beta_2} & \dfrac{1}{(\alpha_1-\bar\beta_2)^2}\\
\end{array}\right| , \nonumber\\ &&g(\alpha_n)=r_n - z + \alpha_n.
\label{m_f} \eea Note that the partial derivatives with respect to
$\a_1$ in (\ref{m_f}) and (\ref{ernst_pot}), besides acting on the
explicit expressions of $\a_1$, also act on $r_1$ containing
$\a_1$ via (\ref{ernst_pot}).

Formulae (\ref{ernst_pot})--(\ref{m_f}) fully describe the
``extreme--non-extreme'' double--Kerr systems which can be divided
into two major groups: ($a$) binary systems composed of one
extreme and one subextreme Kerr particles (see figure~1(a)) and
($b$) configurations composed of one extreme and one superextreme
Kerr particles (figure~1(b)). The position of the extreme
constituent on the $z$--axis is defined by $\a_1$; the part of the
symmetry axis between the real--valued $\a_2$ and $\a_3$ represent
the location of the subextreme constituent, whereas the cut
joining the complex conjugate $\a_2$ and $\a_3$ define a
superextreme Kerr particle. To tackle the case of two separated
objects, in what follows we assign, without loss of generality,
the following order to $\a_n$: \be \a_1>{\rm Re}(\a_2)\ge{\rm
Re}(\a_3). \label{order} \ee

\begin{figure}[htb]
\centerline{\epsfysize=70mm\epsffile{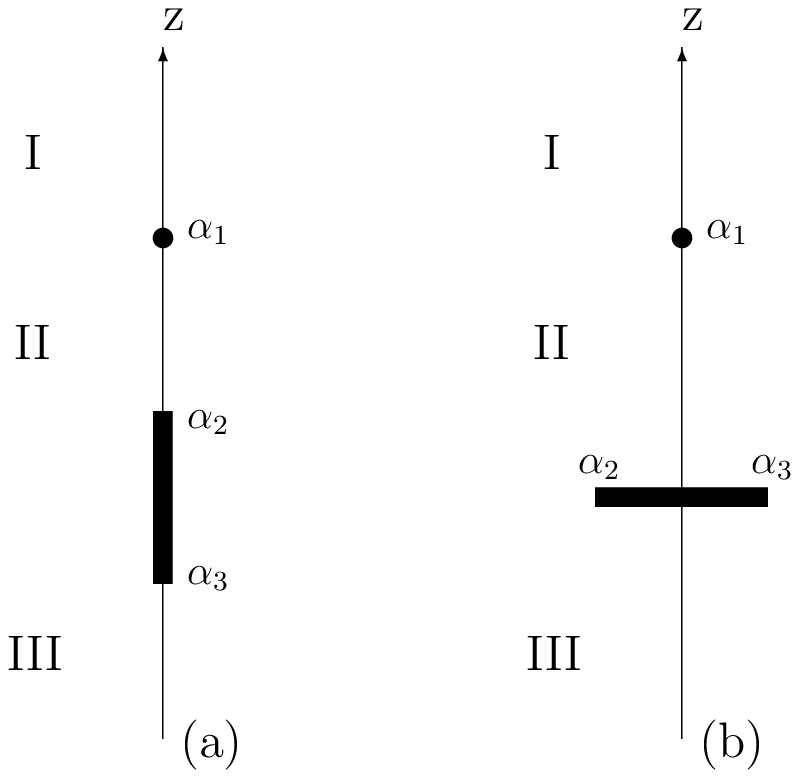}} \caption{Two types
of the `extreme--non-extreme' configurations: (a) the
`extreme--subextreme' case, and (b) the `extreme--superextreme'
case.}
\end{figure}

The equilibrium of two constituents due to balance of the
gravitational attraction and spin--spin repulsion forces implies
the regularity of the symmetry axis outside the location of these
constituents, i.e. regularity of the regions I, II and III shown
in figure~1 which correspond to the parts $z>\a_1$, ${\rm
Re}(\a_2)<z<\a_1$ and $z<{\rm Re}(\a_3)$ of the symmetry axis,
respectively. As is well--known \cite{TKi,DHo}, the regularity of
the regions I, II, III will be achieved if the metric functions
$\gamma$ and $\omega$ are equal to zero there: \be
\gamma^{(I,II,III)}=0, \quad \omega^{(I,II,III)}=0.
\label{eq_cond} \ee

Conditions $\gamma^{(I)}=0$, $\gamma^{(III)}=0$, $\omega^{(I)}=0$
are satisfied automatically by construction of formulae
(\ref{m_f}). The fulfilment of the condition $\omega^{(III)}=0$ is
equivalent to the requirement of asymptotic flatness of the metric
under consideration. Taking into account that on the upper part of
the symmetry axis ($z>\a_1$) the potential $\E$ assumes the form
\bea &&\E(0,z)=1+\frac{e_1}{z-\beta_1}+\frac{e_2}{z-\beta_2},
\nonumber\\
&&e_1=\frac{2(\beta_1-\a_1)^2(\beta_1-\a_2)(\beta_1-\a_3)}
{(\beta_1-\bar\beta_1)(\beta_1-\beta_2)(\beta_1-\bar\beta_2)},
\nonumber\\
&&e_2=\frac{2(\beta_2-\a_1)^2(\beta_2-\a_2)(\beta_2-\a_3)}
{(\beta_2-\beta_1)(\beta_2-\bar\beta_1)(\beta_2-\bar\beta_2)},
\label{e_axis} \eea the asymptotic flatness condition (vanishing
of the angular momentum mono\-pole moment) reduces to \be {\rm
Im}(e_1+e_2)=0. \label{flat_cond} \ee

Therefore, the latter equation together with the remaining two
conditions \be \gamma^{(II)}=0 \quad \mbox{and} \quad
\omega^{(II)}=0 \label{cond_g2} \ee constitute the entire system
of the balance equations whose solutions define equilibrium
configurations of the extreme and non--extreme Kerr particles.

We now turn to concrete examples.

%%%%%%%%%%%%%%%%%%%%%%%%%%%%%%%
\section{Particular equilibrium states}
%%%%%%%%%%%%%%%%%%%%%%%%%%%%%%%

The approach which eventually proved to be successful for solving
numerically the algebraic system of equations
(\ref{flat_cond})--(\ref{cond_g2}) is essentially based on the
following two ingredients: ($i$) the fact that the metric
coefficients $\gamma$ and $\omega$ of the double--Kerr solution
are step functions on the symmetry axis \cite{TKi}, so that in
particular they take constant values on the part of the $z$--axis
between the particles, and ($ii$) a fortunate concrete choice of
the unknowns in the balance system for fixing the remaining
parameters.

With regard to ($i$) one can show, after expanding determinants
with the aid of Laplace's rule, that equation $\gamma^{(II)}=0$
reduces to solving the equation \be {\rm Re}(D)=0, \quad
D=\left|\begin{array}{cc} \dfrac{1}{\alpha_2 - \bar\beta_1} &
\dfrac{1}{\alpha_3 - \bar\beta_1} \vspace{0.15cm}\\
\dfrac{1}{\alpha_2 - \bar\beta_2}
& \dfrac{1}{\alpha_3 - \bar\beta_2}\\
\end{array}\right| \cdot \left|\begin{array}{cc} \dfrac{1}{\alpha_1 - \beta_1}
& \dfrac{1}{(\alpha_1 - \beta_1)^2}
\vspace{0.15cm}\\\dfrac{1}{\alpha_1 - \beta_2}
& \dfrac{1}{(\alpha_1 - \beta_2)^2}\\
\end{array}\right|, \label{eq_g1} \ee if $\a_2$ and $\a_3$ are real, and
equation \be {\rm Im}(D)=0, \label{eq_g2} \ee if $\bar\a_3=\a_2$.
The easiest practical way of obtaining the explicit form of
$\omega^{(II)}$ is to consider the coefficient at the leading
power of $z$ after substituting appropriately $r_n$ by $|z-\a_n|$
in the expression of $\omega$, but we do not write down here the
resulting expression because of its cumbersome form. What ($ii$)
is concerned, one has to remember that the system
(\ref{flat_cond})--(\ref{cond_g2}) is underdetermined (three
equations for seven unknowns $\a_1$, $\a_2$, $\a_3$, ${\rm
Re}(\beta_1)$, ${\rm Im}(\beta_1)$, ${\rm Re}(\beta_2)$, ${\rm
Im}(\beta_2)$). To get the numerical roots, we were fixing the
values of $\a_1$, $\a_2$, $\a_3$ and ${\rm Re}(\beta_1)$, then
searching for ${\rm Im}(\beta_1)$, ${\rm Re}(\beta_2)$ and ${\rm
Im}(\beta_2)$.

The next step after finding numerical values of $\a$'s and
$\beta$'s which define an equilibrium configuration is the
estimation of individual masses and angular momenta of the
constituents with the aid of Komar integrals \cite{Kom}. In the
case of two subextreme Kerr constituents this can be easily done
via very concise Tomimatsu's formulae \cite{Tom}; however, in the
presence of an extreme object which can be accompanied by a
superextreme Kerr particle the integration should be carried out
over the surface of a cylinder surrounding one or another
constituent, and the explicit formulae for the individual masses
and $M_i$ and angular momenta $J_i$ of the particles are
\cite{MRS}
\begin{eqnarray}
M_i &=&\frac{1}{4} \left(\int_{z_d}^{z_u}[\rho (\ln f),_\rho -
\omega \Omega,_z]_{\rho = \rho_0} \d z + \int_{0}^{\rho_0} [\rho
(\ln f),_z + \omega \Omega,_\rho]_{z = z_u}
\d\rho\right.\nonumber\\ &-& \left. \int_{0}^{\rho_0} [\rho (\ln
f),_z + \omega \Omega,_\rho]_{z = z_d} \d\rho\right) ,
\nonumber\\
J_i &=&-\frac{1}{8} \left(\int_{z_d}^{z_u} [2 \omega - 2 \rho
\omega (\ln f),_\rho + (\rho^2 f^{-2} + \omega^2 )\Omega,_z]_{\rho
= \rho_0} \d z\right.\nonumber\\  &-& \int_{0}^{\rho_0} [ 2 \rho
\omega (\ln f),_z + (\rho^2 f^{-2} + \omega^2 )\Omega,_\rho]_{z
=
z_u} \d\rho\nonumber\\
&+&\left. \int_{0}^{\rho_0} [ 2 \rho \omega (\ln f),_z + (\rho^2
f^{-2} + \omega^2 )\Omega,_\rho]_{z = z_d} \d\rho\right),
\label{mj_kom}
\end{eqnarray} $z_u$ and $z_d$ denoting locations on the $z$--axis
of the centers of cylinder's upper and lower bases, $\rho_0$ being
the radius of the bases. Mention that the sum of individual masses
and the sum of individual angular momenta should coincide,
respectively, with the total mass $M$ and total angular momentum
$J$ calculated from the asymptotic form of the potential
(\ref{ernst_pot}): \be M = -\frac{1}{2} {\rm Re}(e_1 + e_2), \quad
J = \frac{1}{2}{\rm Im}(e_1 \bar\beta_1+e_2\bar\beta_2).
\label{mj_tot} \ee

It is highly important for the total and both individual masses to
assume positive values because only in this case the corresponding
equilibrium configuration can be considered physically acceptable.
The particular equilibrium states obtained by us can be described
as follows.

\medskip

{\it Extreme--subextreme equilibrium configurations}. The general
characteristic feature of all such configurations is that they
necessarily involve at least one constituent with negative mass,
which is in line with the Manko--Ruiz theorem \cite{MRu1} on the
absence of two--Kerr black hole equilibrium states with positive
Komar masses. Equilibrium is possible between two constituents
with negative masses, and also when the mass of either the extreme
or the subextreme particle is negative. In table~1 one finds
particular $\a$'s and $\beta$'s determining three
extreme--subextreme equilibrium configurations (the approximate
numerical values are given up to three decimal places), and in
table~2 the corresponding physical characteristics of the
constituents are shown (index~1 refers to the upper, extreme
constituent, and index~2 to the lower, subextreme one).

\begin{table}
\caption{Particular numerical solutions of the balance equations
in the `extreme--subextreme' case.} \vspace{0.5cm}
\begin{tabular}{ccccc} \hline $\alpha_1$ & $\alpha_2$ &
$\alpha_3$ & $\beta_1$ & $\beta_2$
\\ \hline $2.4$ & $-0.3$ & $-1.1$ & $0.3 + 6.061 i$ & $-3.661 +
2.1 i$\\ $3.1$ & $-0.1$ & $-1.3$ & $-1.4 + 0.071 i$ & $5.165 +
2.364 i$\\ $2.7$ & $-0.1$ & $-0.9$ & $0.24 + 0.996 i$ & $3.451 +
0.713 i$\\ \hline
\end{tabular}
\end{table}

\begin{table}
\caption{Masses and angular momenta in the `extreme--subextreme'
equilibrium configurations from table~1.} \vspace{0.5cm}
\begin{tabular}{cccccc} \hline $M_1$ & $J_1$ & $M_2$ &
$J_2$ & $M$ & $J$
\\ \hline $13.499$ & $89.597$ & $-8.438$ & $-68.864$ & $5.061$ &
$20.733$\\ $-2.699$ & $-7.089$ & $1.334$ & $0.79$ & $-1.365$ & $-6.299$\\
$-0.673$ & $-0.375$ & $-0.818$ & $-0.52$ & $-1.491$ & $-0.895$\\
\hline
\end{tabular}
\end{table}

In figure~2 we have plotted the stationary limit surfaces ($f=0$)
for the above equilibrium states, and one can see that the
constituents with negative masses develop the massless ring
singularities located on the $f=0$ surface.

\begin{figure}[htb]
\centerline{\epsfysize=50mm\epsffile{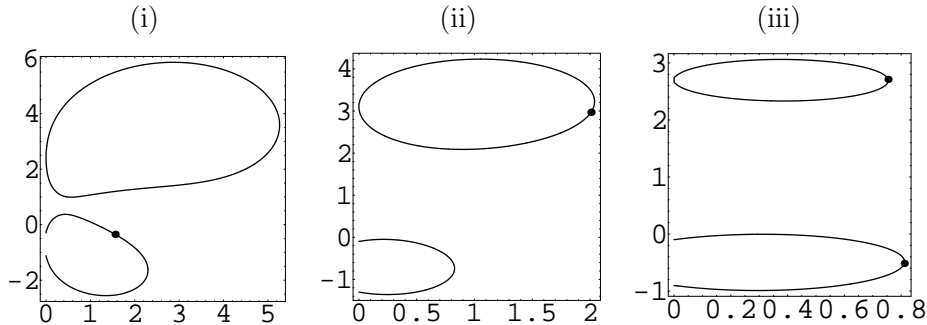}} \caption{Stationary
limit surfaces in the `extreme--subextreme' equilibrium
configurations from table~1.}
\end{figure}

\medskip

{\it Extreme--superextreme equilibrium configurations}. In this
kind of binary systems equilibrium can take place when ($a$) both
constituents have negative masses, ($b$) one constituent has
negative mass and the other has positive mass, or ($c$) both Kerr
particles possess positive Komar masses. Of course, only the
latter case represents physical interest, so we shall restrict
further consideration exclusively to it.

\begin{table}
\caption{Particular numerical solutions of the balance equations
in the `extreme--superextreme' case.} \vspace{0.5cm}
\begin{tabular}{cccc} \hline $\alpha_1$ &
$\alpha_2=\bar\alpha_3$ & $\beta_1$ & $\beta_2$
\\ \hline $3.9$ & $-1.9-6.45i$ & $-3.7 + 7.6 i$ & $0.312 +
3.588 i$\\ $4.1$ & $-0.8-6.7i$ & $-3.8 + 7.9 i$ & $2.68 + 1.42 i$\\
$3.6$ & $-6.3i$ & $-3.9 + 7.5 i$ & $3.43 + 0.173 i$\\
$3.4$ & $-2.1-5.8i$ & $-3.05 + 6.45 i$ & $-0.316 + 3.716 i$\\
\hline
\end{tabular}
\end{table}

\begin{table}
\caption{Masses and angular momenta in the `extreme--superextreme'
equilibrium configurations from table~3.} \vspace{0.5cm}
\begin{tabular}{cccccc} \hline $M_1$ & $J_1$ & $M_2$ &
$J_2$ & $M$ & $J$ \\
\hline $4.204$ & $19.767$ & $1.184$ & $13.247$ & $5.388$ & $33.014$\\
$1.6$ & $3.653$ & $2.82$ & $26.323$ & $4.42$ & $29.976$\\
$0.178$ & $0.064$ & $3.895$ & $29.891$ & $4.073$ & $29.955$\\
$4.126$ & $17.982$ & $0.54$ & $5.483$ & $4.666$ & $23.465$\\
\hline
\end{tabular}
\end{table}

In table~3 four different numerical solutions of the system
(\ref{flat_cond})--(\ref{cond_g2}) are given, and in table~4 the
corresponding masses and angular momenta calculated with the aid
of the formulae (\ref{mj_kom}), (\ref{mj_tot}) are shown. In
figure~3 we have plotted the stationary limit surfaces for the
equilibrium states from table~3, and one can see that because of
the positiveness of the masses of the extreme and superextreme
constituents no massless ring singularities appear on these
surfaces.

\begin{figure}[htb]
\centerline{\epsfysize=100mm\epsffile{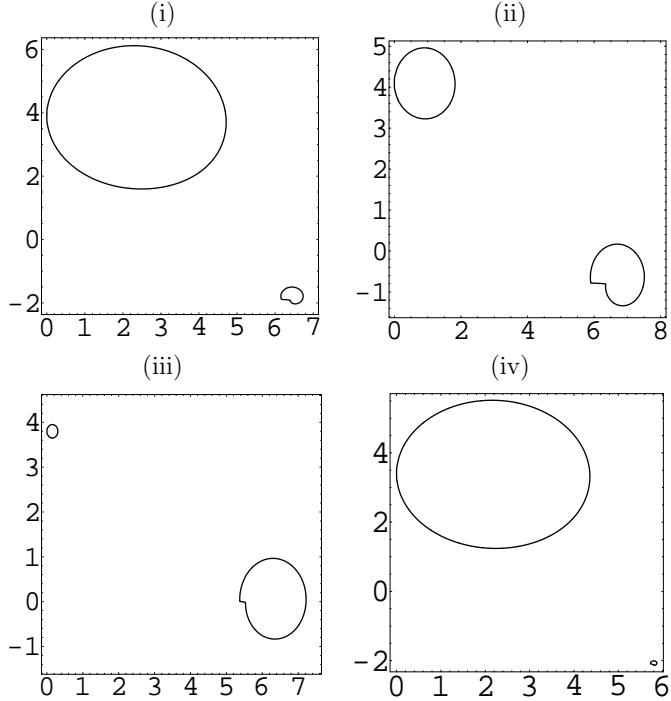}}
\caption{Stationary limit surfaces in the `extreme--superextreme'
equilibrium configurations from table~3.}
\end{figure}

%%%%%%%%%%%%%%%%%%%%%%
\section{Discussion}
%%%%%%%%%%%%%%%%%%%%%%

Probably the physically most interesting result following from our
investigation is that the well--known relation $|J|=M^2$ valid for
a single extreme Kerr black hole \cite{Ker} does not necessarily
hold in a binary system of Kerr particles involving an extreme
constituent. The data from tables~2 and 4 provide evidence that
the extreme Kerr constituent in balance with the other
non--extreme one is characterized by the inequality \be
|J_1|<M_1^2 \label{mj_extr1} \ee in the `extreme--subextreme'
configurations, and by the inequality \be |J_1|>M_1^2
\label{mj_extr2} \ee in the `extreme--superextreme'
configurations.

It should be pointed out that all the `extreme--non-extreme'
equilibrium configurations considered verify the general
equilibrium law of two Kerr particles established by Manko and
Ruiz \cite{MRu2}: \be
J\pm(M+s)^2+s\left(\frac{J_1}{M_1}+\frac{J_2}{M_2}\right)=0,
\label{eq_law} \ee where $s$ is the coordinate distance between
the particles, i.e., in our case $s=(2\a_1-\a_2-\a_3)/2$. This may
be viewed as an independent confirmation of the correctness of our
results.

The present paper, therefore, complements the study of equilibrium
configurations in the extended double--Kerr solution and provides
counter--examples to the conjecture about the non--existence of
`extreme--non-extreme' equilibrium states put forward in
\cite{MRS}. In view of the results obtained we may speculate that
most probably the equilibrium states involving one extreme object
can also arise in the double--Kerr--Newman systems; however,
finding concrete equilibrium configurations of that kind remains a
task for the future.

\vspace{1cm}

\noindent{\bf Acknowledgements}

\vspace{.5cm}

\noindent VSM would like to thank the Department of Fundamental
Physics of the Salamanca University for its kind hospitality and
financial support of his visit, JDSG acknowledges financial
support from COLCIENCIAS of Colombia. This work was also supported
by Project 45946--F (CONACyT, Mexico), Project BFM2003--02121
(Ministerio de Ciencia y Tecnolog\'\i a, Spain) and by
Project~5116 (DIF de Ciencias of the Santander Industrial
University, Colombia).

\end{document}